# Brightness of the Qianfan Satellites

Anthony Mallama[1], Richard E. Cole[1], Bram Dorreman[2], Scott Harrington and Nick James

[1] IAU Centre for the Protection of Dark and Quiet Skies from Satellite Constellation Interference
[2] Belgian Working Group Satellites

2024 September 30

Correspondence: anthony.mallama@gmail.com

Abstract

Observed magnitudes of Qianfan spacecraft range from 4 when they are near zenith to 8 when low in the sky. Nearly all of the observations can be modeled with a nadir-facing flat antenna panel and the underside of a zenith-facing solar array, both with Lambertian reflectance properties. These satellites will impact astronomical research unless their brightness is reduced.

## 1. Introduction

The first 18 spacecraft of the Qianfan mega-constellation were launched by Shanghai SatCom Satellite Technology on 2024 August 6. When fully populated Qianfan will add 14,000 new communication satellites to low Earth orbit.

This initial batch of spacecraft are in polar orbits inclined 89 degrees to the Earth's equator. Their altitudes near 800 km are intermediate in height between the lower Starlink satellites and the higher OneWeb constellation.

This paper reports on brightness measurements of Qianfan spacecraft recorded between August 12 and September 9. Section 2 describes the observations. Section 3 characterizes the satellites' brightness empirically, while Section 4 models that brightness physically. Section 5 discusses the impact on astronomical research and Section 6 summarizes our findings.

## 2. Observations

The Qianfan satellite train was spotted with the unaided eye and recorded on wide-field video even before accurate orbital elements became available. One frame of a higher-resolution video shows the train passing near zenith in Figure 1.

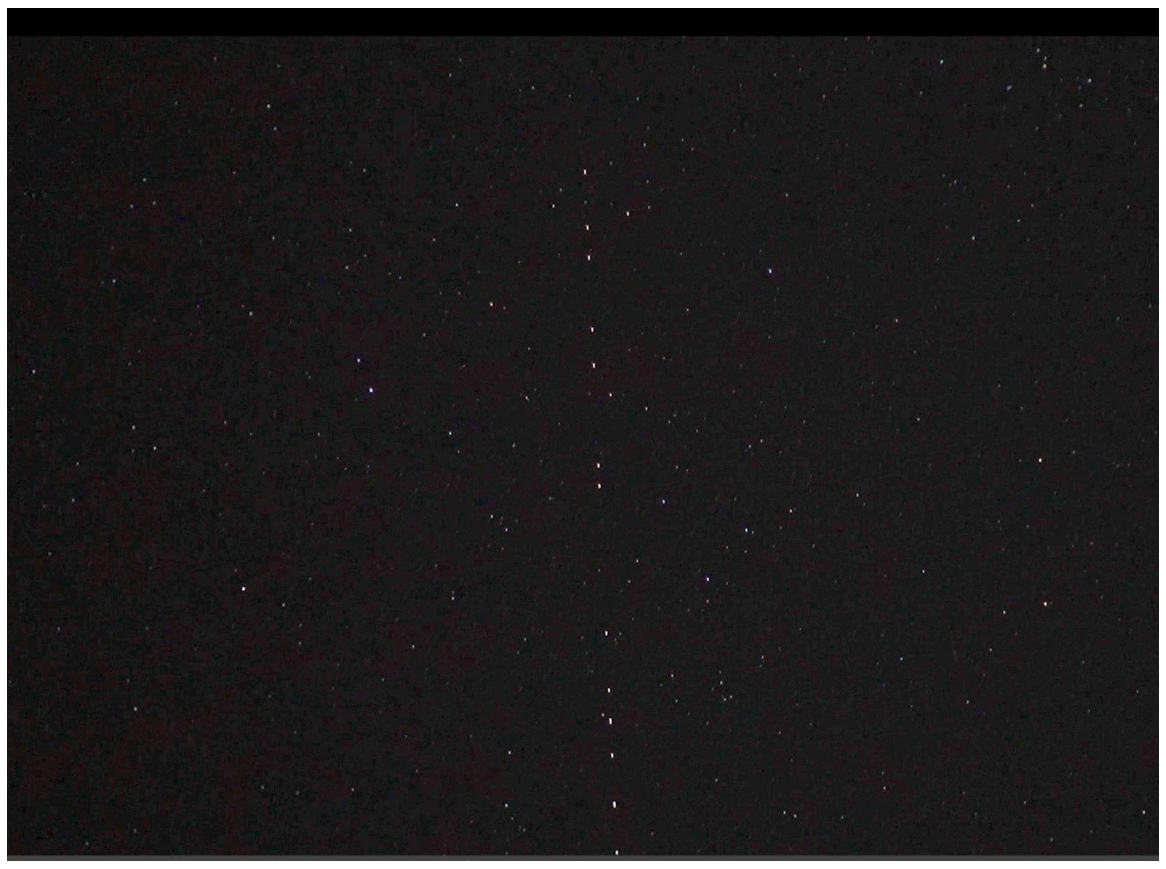

*Figure 1. Detail of a frame from the higher resolution video of 2024 August 12.*

Visual observers began recording brightness after ephemeris data were published. Magnitudes are determined by comparing the satellites to nearby reference stars. The angular proximity between satellites and stellar objects accounts for variations in sky transparency and sky brightness. Mallama (2022) describes this method in more detail. Magnitudes of three satellites in the video associated with Figure 1 were determined in a similar manner.

3. Empirical brightness characterization

The Qianfan satellites appear brighter near zenith than at low elevations as shown in Figure 2. This behavior is not unexpected because high elevations correspond to smaller ranges where the inverse square law of light makes the spacecraft appear more luminous. The least square fit is,

$$Y = 8.613 - 0.1086 * X + 0.0006821 * X^2$$

Equation 1

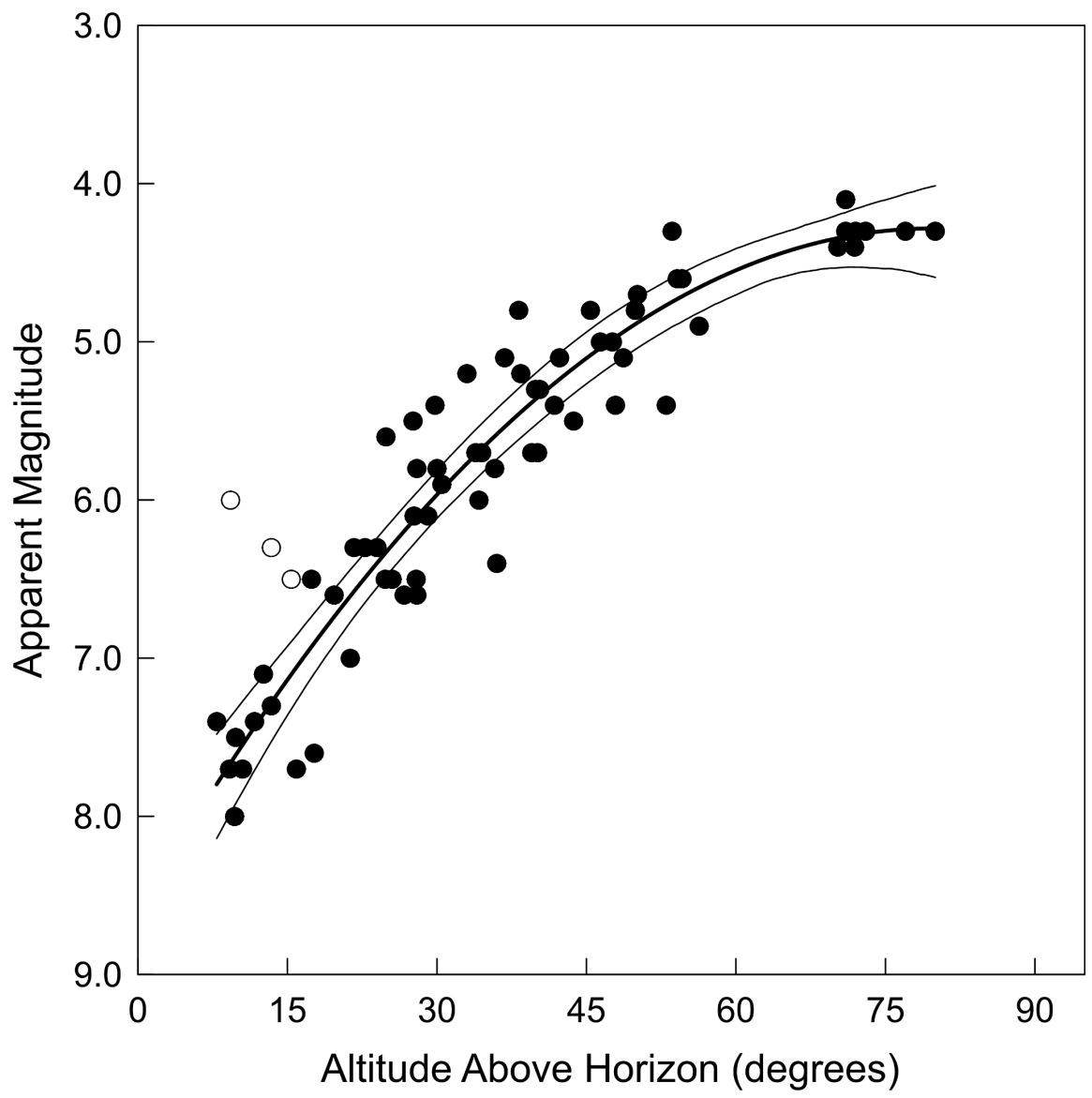

*Figure 2. Apparent magnitude as a function of altitude above the horizon. Thin lines represent the confidence interval. Magnitudes indicated by open circles are discussed in Section 4.*

However, magnitudes do not correlate strongly with range and elevation in the sky for all satellites. Figure 3 demonstrates that the brightness of Starlink satellites is only a weak function of range and that it does not even decrease monotonically with range. This behavior is due to the shape of Starlink satellites and to their attitude. SpaceX adjusts the attitude to make them fainter when the spacecraft are high in the sky.

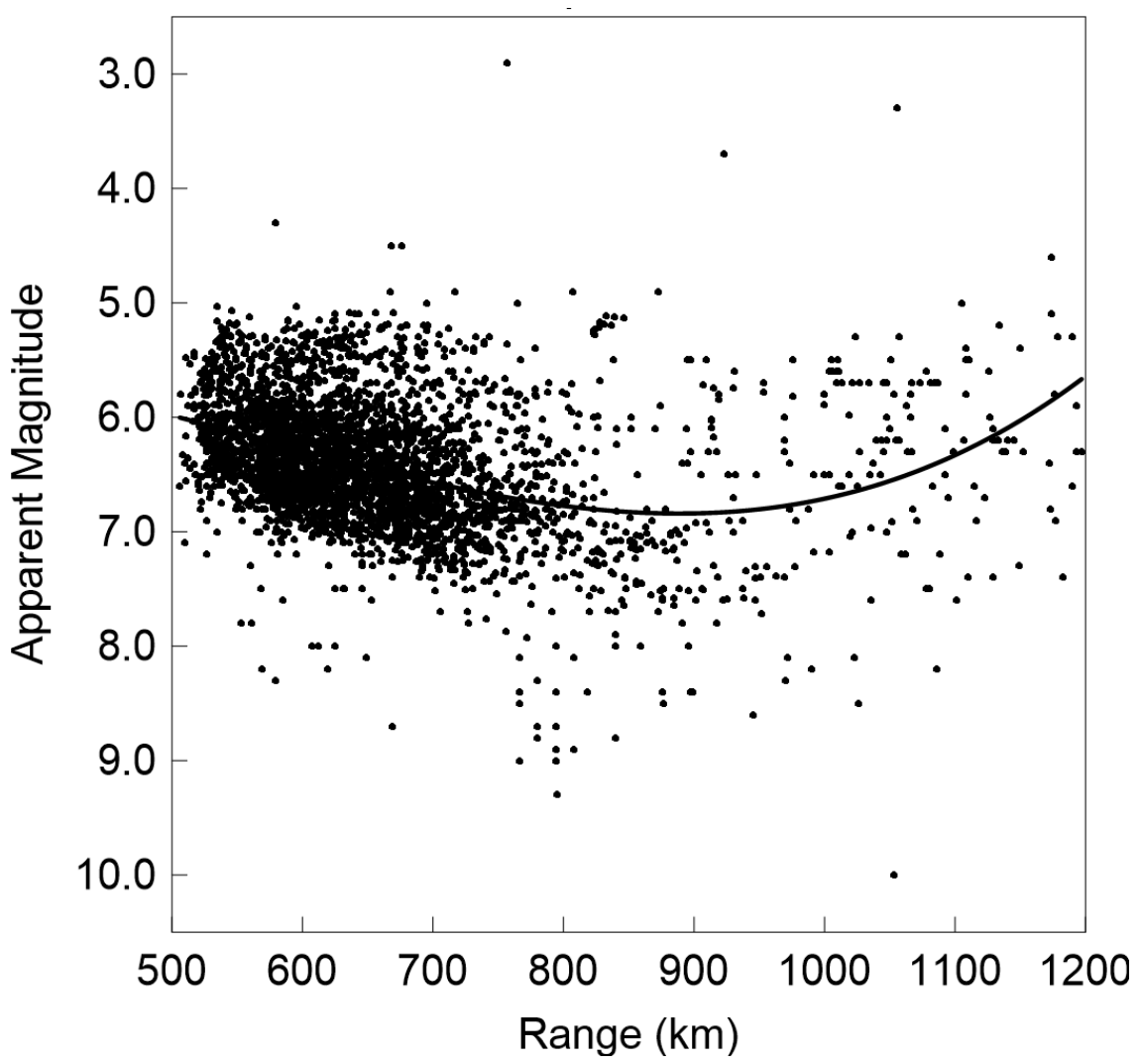

*Figure 3. Starlink brightness is not strongly dependent on range.*

The illumination phase function is another form of brightness characterization; it accounts for varying geometries involving the Sun, satellite and observer. Phase angle is that arc measured at the satellite between directions to the Sun and the observer. Apparent magnitude is adjusted to a uniform distance of 1000 km by applying the inverse square law.

The phase function for Qianfan shown in Figure 4 indicates that they are brighter when viewed at smaller phase angles. In that geometry the satellites and the Sun are in nearly opposite directions as seen by the observer. Thus, the spacecraft are well lit and they appear more luminous. The least squares fit is,

$$Y = 4.103 + 0.01047 * X,$$

Equation 2

While empirical characterization is useful as a tool for predicting satellite brightness, the physical modeling described in the next section provides a deeper understanding.

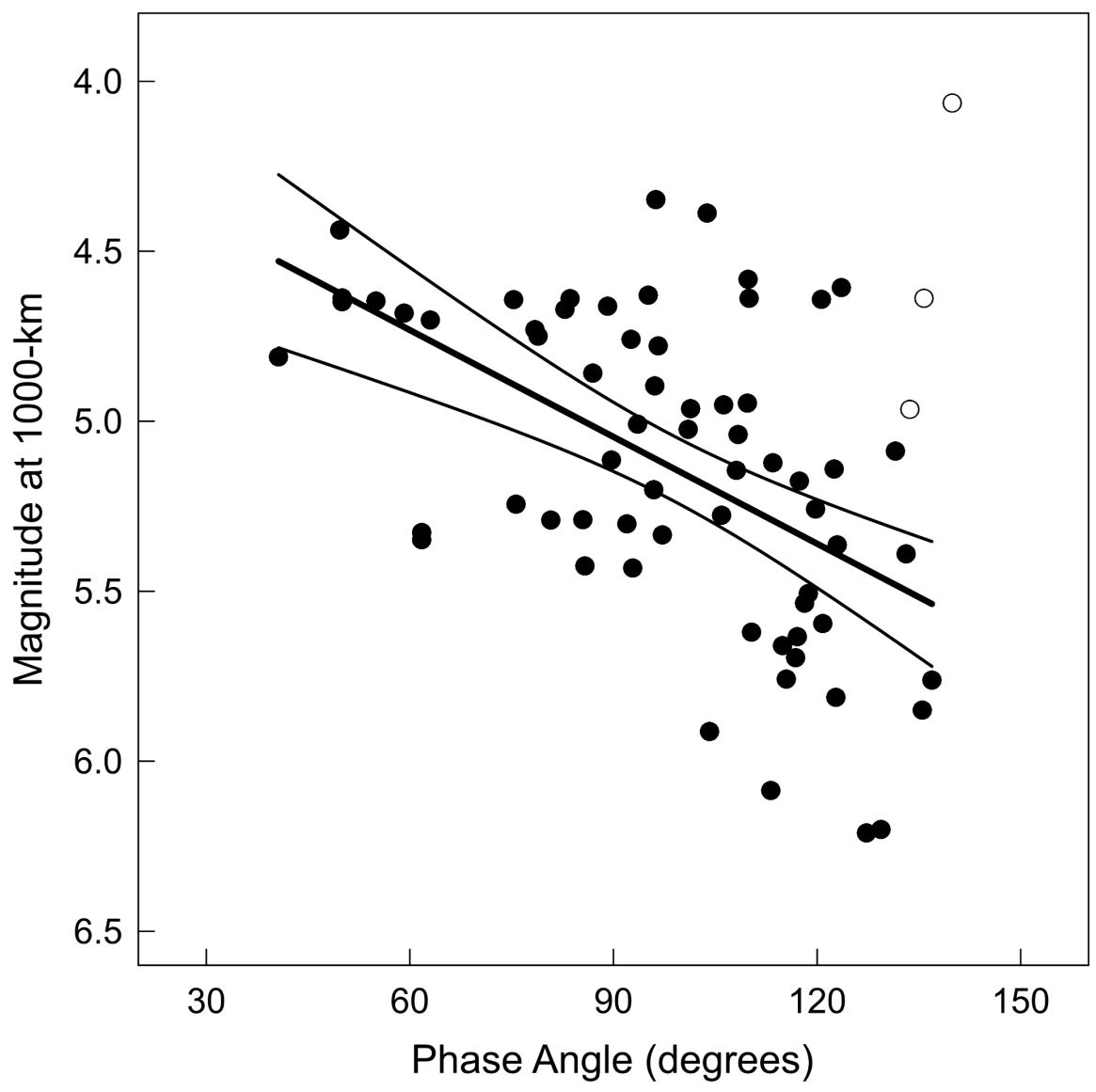

*Figure 4. A linear fit to the phase function indicates brighter magnitudes at smaller angles. Magnitudes indicated by open circles are discussed in Section 4.*

## 4. Physical modeling

Simple physical brightness models have been developed for previous spacecraft such as Starlink (Cole, 2021). In these models the satellite is represented by a small number of surfaces and the reflection from these surfaces calculated using the position of the Sun and satellite with respect to the observer. The size and orientation of the surfaces and the type of reflection from those surfaces (for example, diffuse or specular) is adjusted depending on our knowledge of a particular spacecraft's design. Speculation about the design and estimation of surface properties is sometimes required when satellite operators do not disclose all the details.

This approach has been successful in accurately predicting observed brightness while also confirming aspects of spacecraft design and operation that had not been published. For example, several temporary fadings of Bluewalker 3 were attributed to tilting the solar array towards the Sun, and the model allowed measurement of that tilt angle (Mallama et al, 2023).

The same approach was utilized on the Qianfan observations obtained to date. Images of a Qianfan spacecraft design shown in Figure 5 suggest a single solar panel that is only articulated to rotate around the horizontal axis. As with Starlink spacecraft, we suspect that the base of the Qianfan satellites with their communication antennas would be Earth-facing during operations.

The long axis of the solar panel could be oriented either at right-angles to the velocity vector, parallel to velocity vector, or at right-angles to the Sun azimuth. Each of these modes might have advantages for tracking the Sun in various positions with respect to the orbit of the spacecraft. The Qianfan spacecraft were injected into an orbit where the Sun would initially pass through the zenith on each revolution.

For the purposes of modeling their brightness, the Qianfan spacecraft are taken to be a surface representing the antenna panel and potentially a second surface representing the solar panel, differently oriented to the antenna panel.

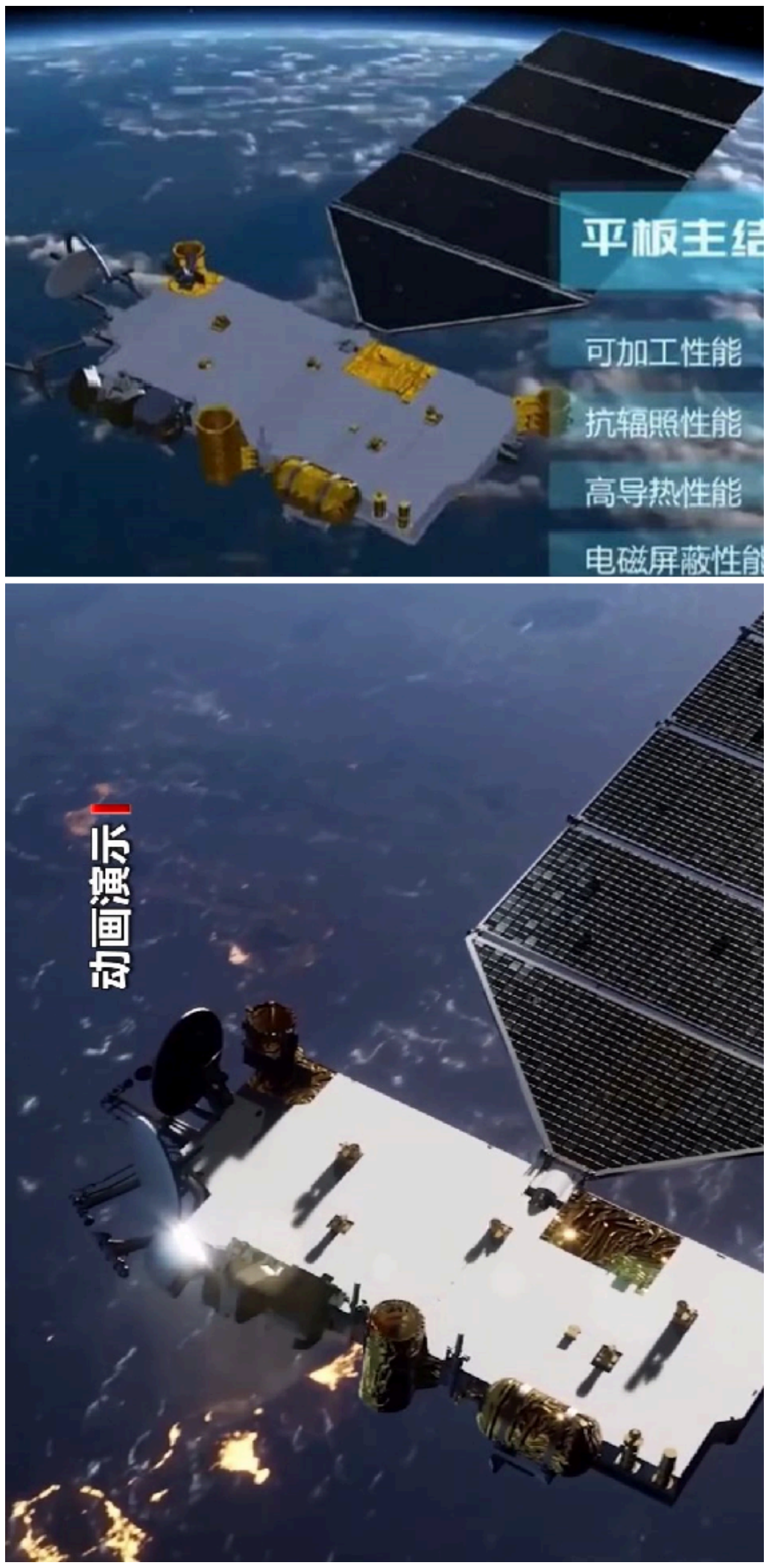

*Figure 5. Two published images of a Qianfan spacecraft. Credit:* [*https://www.youtube.com/watch?v=IG9VDSi8990*](https://www.youtube.com/watch?v=IG9VDSi8990)

The Starlink spacecraft have mirrored surfaces on their Earth-facing panels to direct sunlight into space rather than allowing it to scatter toward the ground. However, as there is no published evidence that the Qianfan spacecraft use this technique, it was assumed that the spacecraft surfaces reflect diffusely, as a so-called Lambertian Surface.

A model was developed using these assumptions and applied to the Qianfan observations. The adjustable parameters are an offset magnitude that represents the size and reflectivity of a single surface and the pointing direction of that surface.

Three cases were considered with the reflecting surface oriented as follows:

a. Earth facing
b. Sun facing
c. Orbital velocity-vector facing

A comparison of the measured magnitudes and the model predictions is shown in Figure 6. As can be seen, only the Earth facing surface case a) has any match to the data. For cases b) and c) most of the predictions are faint because the sunlit face of the surface is pointed away from the observer. Most of the observations were made to the north, close to local midnight, with the Sun illuminating the far side of the surface in the model.

It is clear that an Earth facing surface is making the major contribution to the brightness of the spacecraft.

The majority of the observations are well characterized by the Earth facing model with predictions lying within 0.5 magnitude of the observed values as indicated by the dotted red lines. In general, observations over a range of four magnitudes (a factor of 40 in optical flux) are well modeled by this case.

This indicates that the Qianfan spacecraft are currently oriented with the base of the spacecraft horizontal. Adjustment of the pointing direction of the Earth facing surface by more than 5° makes the model fit to the observations considerably worse.

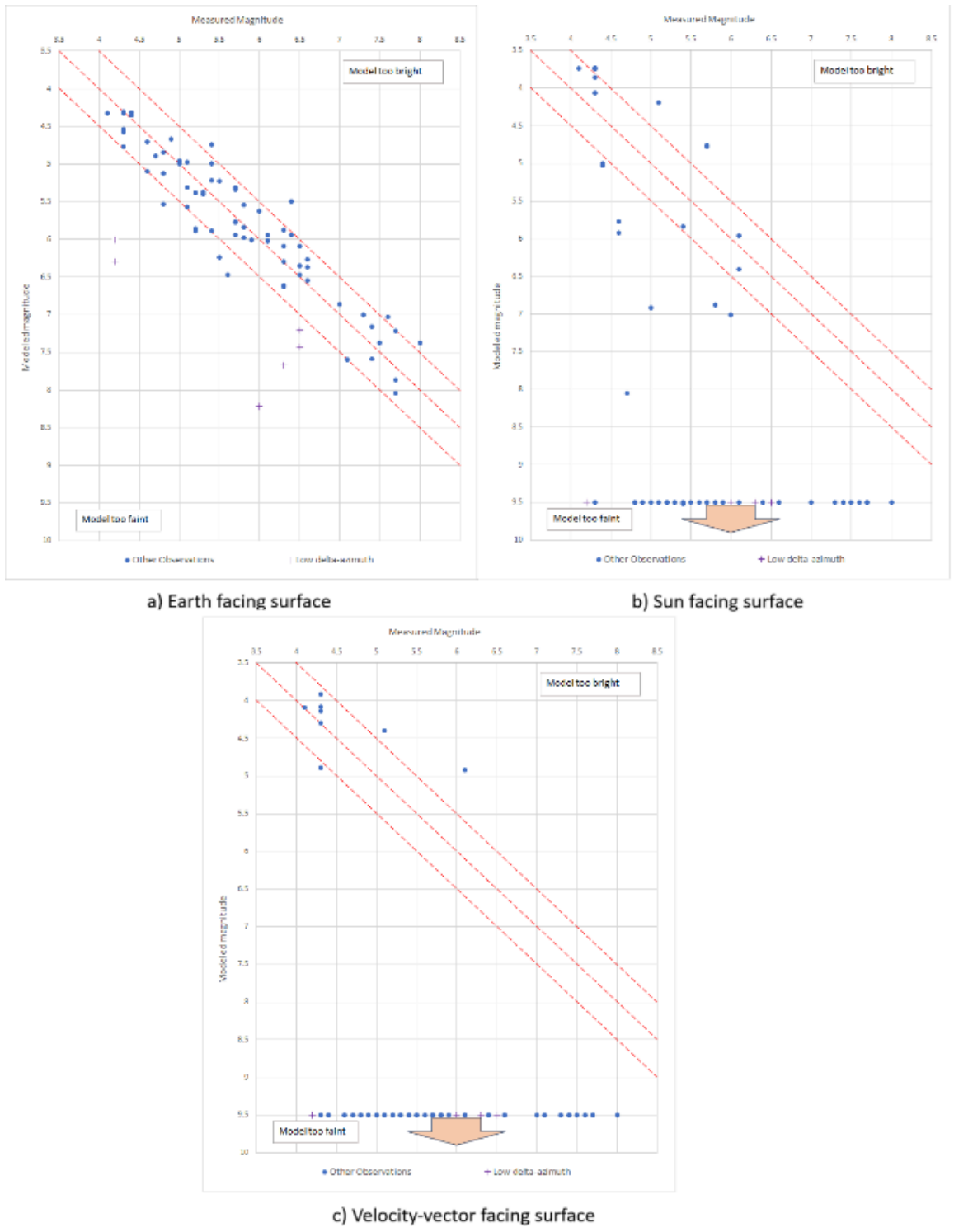

*Figure 6. A comparison of the Qianfan measured and modeled magnitudes, for reflecting surface orientations a, b and c*

To investigate the few observations where the model deviates significantly, a number of parameters were assessed for potential correlation with the errors in the predictions. The best correlation found was with the angle between the observed azimuth of the spacecraft and the azimuth of the Sun (the delta-azimuth), displayed in Figure 7. The deviations occur when the observation is less than 5° from the azimuth of the Sun. This behavior is often seen on other spacecraft in similar positions and can be attributed to forward-scattering of sunlight (rather than diffuse or Lambertian reflection) from Earth-facing surfaces on spacecraft.

This correspondence of the azimuth of Qianfan spacecraft and the Sun was only observed on two passes and further observations can be made when the satellites' orbits evolve away from their orientation at launch.

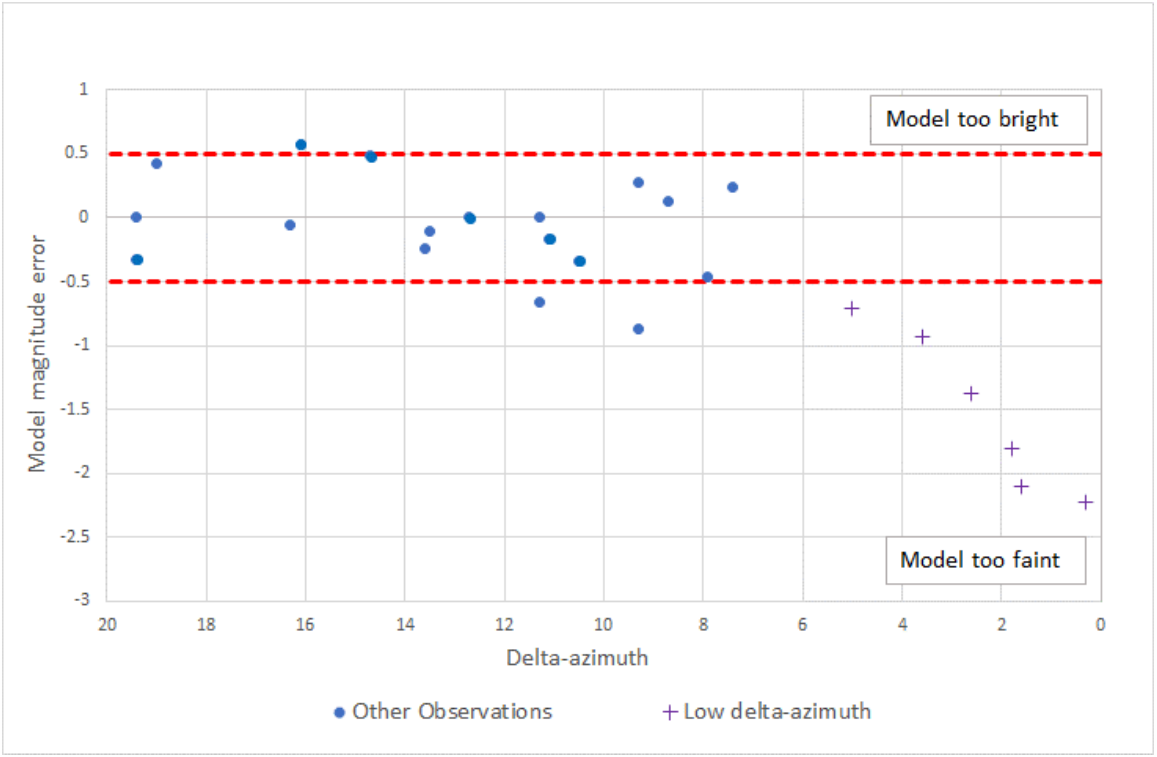

*Figure 7. Model errors plotted against Delta-azimuth, the angle between the satellite and solar azimuths. The model deviates when that angle is small.*

Reflected light from a solar panel differently oriented to the antenna panel could be contributing to the observed brightness in the observations. Given that the available animations of the Qianfan design show the long axis of the solar panel oriented at right angles to the velocity vector, the panel may be rotated around that axis under some Sun conditions to increase power generation. As discussed above (Figure 6b), a large Sun facing panel would make a significant brightness contribution to a subset of the observations but is not evident in the data. As a general conclusion, the current observation set does not indicate a sun-pointing solar array so the array could have been maintained pointing to the zenith during this period, when the Sun was passing close to the zenith on each orbit. Rotation of the solar array will be required under other Sun conditions at the spacecraft, particularly when the Sun never rises to a high elevation.

When the Qianfan orbit evolves so the Sun does not pass through the zenith at the spacecraft, the solar panel may be rotated around a horizontal axis to point towards the Sun to obtain sufficient solar energy. Some change in the brightness of the Qianfan spacecraft will result, depending on the panel's angle of rotation. Observations at that time may indicate whether the Qianfan operators are attempting to minimize the final brightness of the spacecraft by control of the solar panel angle.

While only a relatively few observations have been made, there is no indication of reflection from mirrored surfaces on the spacecraft. At this time, it does not appear the Qianfan spacecraft utilize the technique employed on Starlink satellites to reduce brightness.

## 5. Impact on astronomy

Bright satellites interfere with scientific observations (Barentine et al. 2023) and with aesthetic appreciation of the night sky (Mallama and Young, 2021). Tyson et al. (2020) determined that streaks from satellites brighter than magnitude 7 could not effectively be removed from Rubin Observatory images for their Legacy Survey of Space and Time (LSST) program. Meanwhile, the limit for visual observing is about magnitude 6 because the unaided eye can see brighter objects.

Qianfan satellites are brighter than magnitude 6 except when observed at low elevations in the sky. So, they will adversely impact professional and amateur astronomical activities unless the operators mitigate their brightness.

SpaceX made changes to the design of their Starlink satellites because early observations demonstrated that their Generation 1 spacecraft would impact astronomy. That is our motivation for reporting early results for Qianfan.

Furthermore, the analyses in this paper pertain to the first batch of Qianfan satellites at 800 km altitude. The operational satellites are expected to orbit at 500 and 300 km. Spacecraft at those heights would appear approximately 1 and 2 magnitudes brighter, respectively.

## 6. Conclusions

The brightness of Qianfan spacecraft ranges from magnitude 4 when they are near zenith to 8 when low in the sky. Nearly all of the observations can be modeled with a nadir-facing flat antenna panel and the underside of a zenith-facing solar array, both with Lambertian reflection properties. These satellites will impact astronomical research and aesthetic appreciation of the night sky unless their brightness is mitigated.